\documentclass[reqno]{amsart}
\usepackage{graphicx,amsmath}
\usepackage{epstopdf}
\usepackage{float}

\def\bq{\mathbf q}

\def\re#1{(\ref{#1})}
\def\f{\frac}

\begin{document}

\title{Deviation from the Fourier law in room-temperature heat pulse experiments}

\author{Both, S.$^{1}$, Cz\'el, B.$^{1}$, F\"ul\"op, T.$^{13}$, Gr\'of, Gy.$^{1}$, Gyenis, \'A.$^{1}$, Kov\'acs R.$^{123}$, V\'an P.$^{123}$ and Verh\'as, J.$^{3}$}

\address{
$^1$Department of Energy Engineering, BME, Budapest, Hungary and 
$^2$Department of Theoretical Physics, Wigner Research Centre for Physics,
Institute for Particle and Nuclear Physics, Budapest, Hungary and 
$^3$Montavid Thermodynamic Research Group}

\date{\today}

\begin{abstract}
We report heat pulse experiments at room temperature that cannot be described by Fourier's law. The experimental data is modelled properly by the Guyer--Krumhansl equation, in its over-diffusion regime. The phenomenon is due to conduction channels with differing conductivities, and parallel to the direction of the heat flux.
\end{abstract}
\keywords{Guyer-Krumhansl equation, Jefrreys type equation}
\maketitle

\section{Introduction}

Fourier's law of heat conduction is one of the most important laws of physics in our everyday life. Energy consumption of heating systems, power plants and industrial processes are designed and manufactured with the help of the classic formula, expressing that the heat flux $\bq$ is proportional to the temperature gradient:
\begin{equation}
 \bq = - k \nabla T.
\label{Fourier}\end{equation}
Here, $k$ is the heat conduction coefficient. However, at low temperatures and in small systems Fourier's law is violated. The most important related phenomenon is called second sound, that is when temperature disturbances propagate like damped waves. The  Maxwell--Cattaneo--Vernotte (MCV) equation \cite{Cat48a,MorFes53b,Ver58a} introduces the inertia of heat, adding a time derivative term to the Fourier equation:
\begin{equation}
\tau \dot{\bq}+ \bq = -k \nabla T.
\label{MCV}\end{equation}
The coefficient of the time derivative, $\tau$, is called relaxation time. This equation is the simplest model of the second sound phenomenon observed first in liquid Helium \cite{Pes44a}. Later on, the analysis of the theoretical background \cite{GuyKru66a1} resulted in the observation of second sound also in solid crystals, via properly designed experiments \cite{AckGuy68a,AckOve69a,McNEta70a,NarDyn72a}. Here, the heat pulse technology was crucial for the sensitive detection. This experimental technique is practically important in itself, in the form of the flash method, a standard engineering procedure for determining thermal diffusivity \cite{ParEta61a}. Theoretical aspects of heat pulse experiments have been shortly reviewed in \cite{Van15a}.

The MCV equation is not the last theoretical development, there were many independent suggestions that further generalize the Fourier law, predicting additional terms  on various grounds \cite{JosPre89a,JosPre90a,Cha98a,Str11b,Cim09a,Leb14a}. The two simplest, and most discussed models are the Guyer--Krumhansl equation and the Jeffreys type equation \cite{GuyKru66a1,JosPre89a} the two coinciding in one spatial dimension.

The low-temperature heat pulse measurements in dielectric crystals exploit well 
understood microscopic mechanisms of phonon propagation \cite{DreStr93a}. However, the phenomenological considerations, especially the theory of non-equilibrium thermodynamics, predicts a universal, mechanism-independent background of the MCV equation, where the deviation from local equilibrium is restricted only by the second law of thermodynamics \cite{Gya77a,JouAta92b}. 

These ideas, in particular the universality, motivated several authors for finding  non-Fourier heat conduction at room temperature in heterogeneous materials, too. The first experiments reported positive results \cite{Kam90a,MitEta95a}, but these results were not confirmed later, more properly, the attempts of exact reproduction of these experiments are contradictory \cite{HerBec00a,HerBec00a1,RoeEta03a,ScoEta09a}. 

However, genuine thermodynamic theories, like Extended Irreversible Thermodynamics, predicted universality originally only with respect to the MCV equation. That may be the reason that, in the above-mentioned early experiments, the authors have been looking for qualitative phenomena characteristic only to the MCV equation: delay in the arrival time of the pulse and the corresponding abrupt temperature change. 

The thermodynamic derivation of the MCV equation assumes heat flux dependent entropy density. One can obtain a more complete characterisation of the  deviation from local equilibrium by considering a generalized form of the entropy current density as well \cite{Mul68a,Ver83a}. In this respect, the idea of current multipliers is particularly convenient for solving the entropy inequality \cite{Nyi91a1,Van01a2}. For  heat conduction, one obtains a general theory that incorporates all viable known constitutive relations, including MCV and Guyer--Krumhansl (GK). Moreover, it provides a promising modelling framework to reproduce combined second sound and ballistic heat propagation effects \cite{KovVan15a,KovVan15m1}. The universal extension of the MCV equation motivated us to look for non-MCV type extension of the Fourier law in experimental observations where the wave equation signatures are suppressed. In the last years, we performed an experimental-theoretical research in order to identify suitable qualitative signatures of detecting non-Fourier heat conduction beyond the MCV equation
\cite{CzeEta13p1,CzeEta13p2,VanEta13p1}.

In this paper, we show a simple heat pulse experiment at room temperature on a macroscopic sample, where the heterogeneities result in deviation from Fourier law. However, this deviation cannot be modelled by the MCV equation, rather the observed characteristic non-Fourier phenomenon is typical for the GK equation. 

The paper is organized as follows. In the next section, the experimental background is described. Then the inverse problem is formulated. In the third section we analyse a representative measurement and determine the parameters of both the Fourier and of the GK equations fit to the experiment. Finally, we summarize and discuss the results.

\section{Heat pulse experiments at room temperature}

\begin{figure}[ht]
	\centering
	\includegraphics[width=8.5cm,height=6cm]{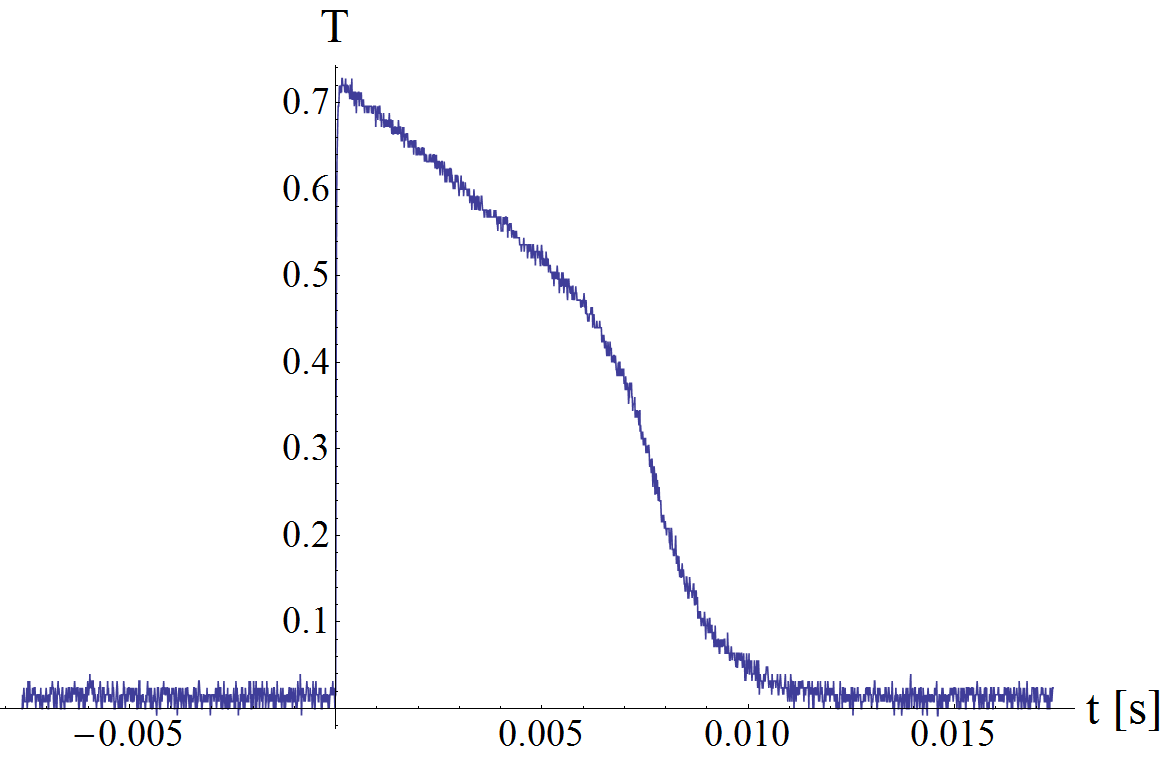}
	\caption{The typical form of the heat pulse, measured by a photovoltaic cell.}
	\label{fig:pulse}
\end{figure}

Our experimental device had been developed at the Budapest University of Technology and Economics, Department of Energy Engineering for industrial use of the flash method, and has been modified for the recent experiments. A flash lamp serves as the heat pulse source at the front end of the sample, and temperature is measured by a pin-thermocouple (K type) at the rear end. The thermocouple and the detector part are insulated from the heat pulse and from the electromagnetic noises. The heat pulse is measured directly at the front end by a photovoltaic cell, providing the triggering signal for the data acquisition. A typical pulse shape can be seen in Figure \ref{fig:pulse}. The electric signal is amplified by a special variable preamplifier and is registered by a digital storage oscilloscope (Tektronix, TDS2014B). 

\begin{figure}[ht]
	\centering
	\includegraphics[width=8.5cm,height=6cm]{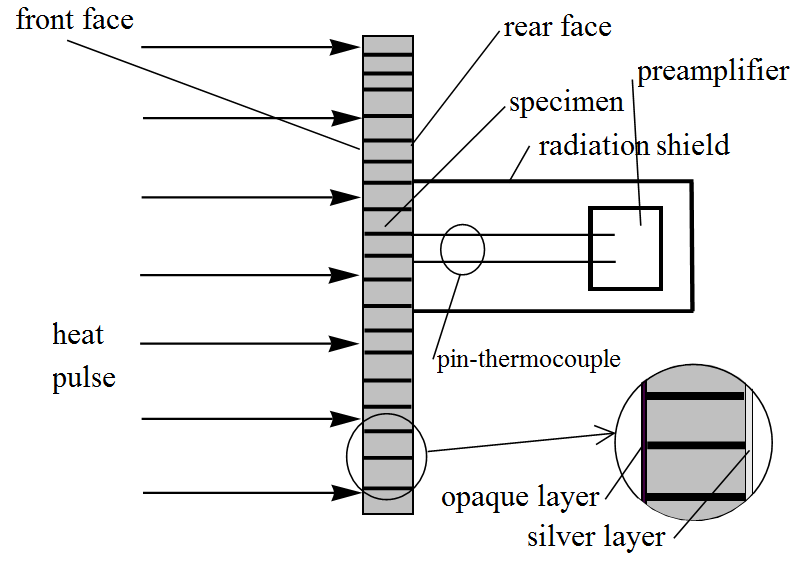}
	\caption{Sketch of the heat pulse experiments.}
	\label{fig:exp}
\end{figure}

\begin{figure}[ht]
	\centering
	\includegraphics[width=10cm]{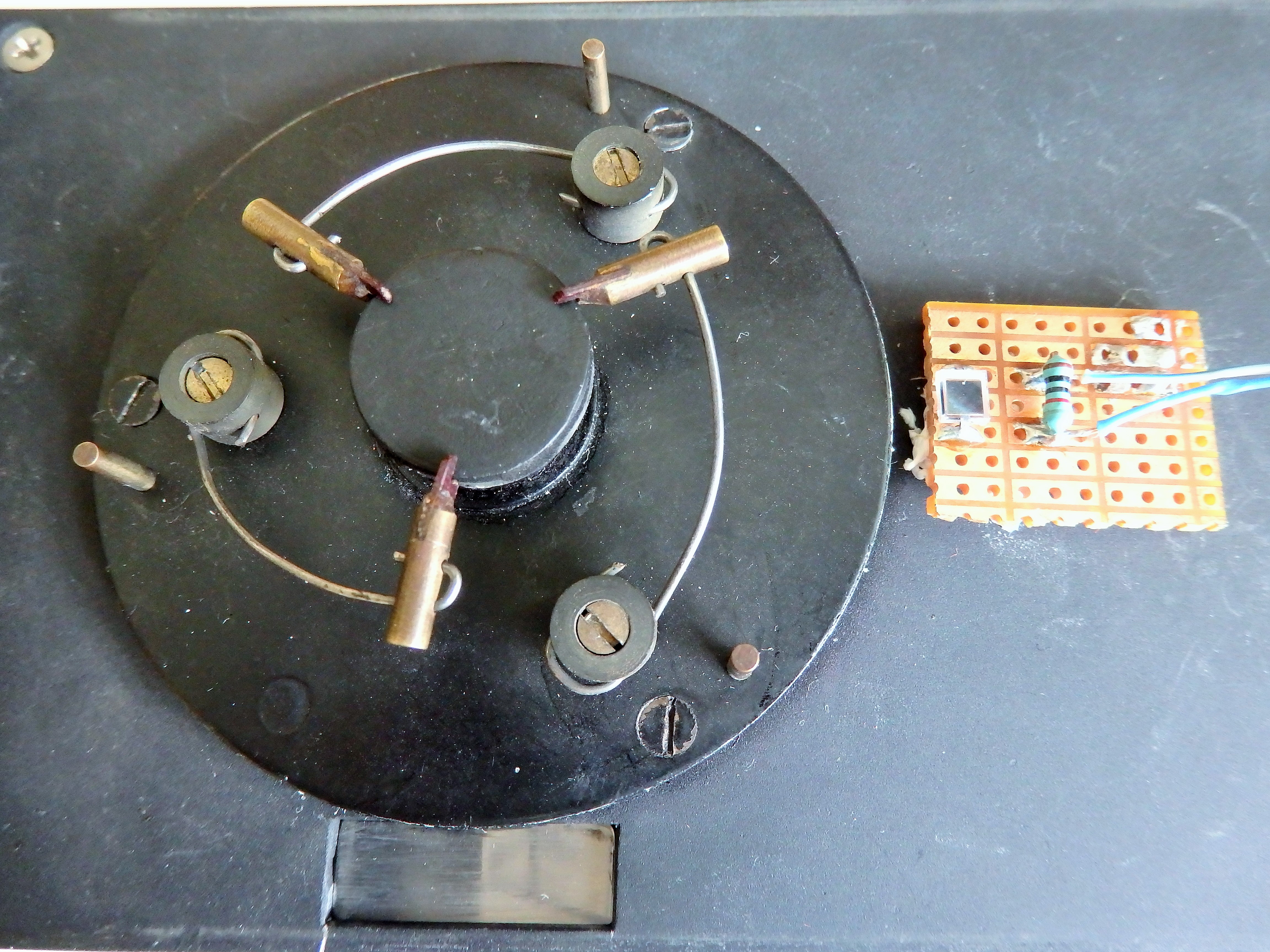}
	\caption{Photo of the sample holder with the inserted specimen. The photovoltaic detector is on the right hand side.}
	\label{fig:samphol}
\end{figure}

A sketch of the experimental setup is shown in Figure \ref{fig:exp}, and a photo of a specimen in the sample holder can be seen in Figure \ref{fig:samphol}. 

The studied specimen in the experiment has a cylindrical shape with $L=3.9 mm$ thickness and $d=19 mm$ diameter. The specimen is composed of aluminium (5 $\mu$m) and polystyrol (15 $\mu$m) layers, arranged parallel to the heat pulse. The front side is painted black to ensure uniform boundary conditions as well as to eliminate the transparency of the sample. At the rear side a silver painting is applied to ensure the thermocouple measures an effective temperature of the heterogeneous material. 

The experimental device was calibrated using several samples with known heat diffusivity. The measurement was performed at $21 ^\circ C$ room temperature.

\section{The Guyer--Krumhansl equation}

We have solved the inverse problem both with the Fourier equation and with the GK equation of heat conduction, and determined the parameters of the differential equations according to the experimental data. The following system of partial differential equations was applied:
\begin{eqnarray}
\rho c \f{\partial T}{\partial t} + \f{\partial q}{\partial x} &=& 0, \label{ebal}\\
\tau \f{\partial q}{\partial t} + q + k \f{\partial T}{\partial x} - l^2 \f{\partial^2 q}{\partial x^2} &=& 0. \label{GK} 
\end{eqnarray}
Here first equation is the balance of internal energy, where $\rho$ is the density, $c$ is the specific heat. $t$ denotes time and $x$ the spatial coordinate in the direction of the heat propagation. The second equation is the GK equation in one dimension, where $k$ is the Fourier heat conduction coefficient, $\tau$ is the relaxation time and $l^2$ is a material parameter of the GK equation, which is nonnegative according to the second law \cite{VanFul12a}, and is expressed with the help of a characteristic length scale $l$. In what follows we solve the above system for the temperature and heat flux fields, $T(x,t)$ and $q(x,t)$, respectively.

Eq. \ref{GK} originally was derived by Guyer and Krumhansl using the linearized Boltzmann equation in order to model low-temperature heat conduction in solids \cite{GuyKru66a1}. Later on it was derived in the framework of kinetic theory by more general assumptions \cite{MulRug98b,DreStr93a}. Nowadays it is extensively researched in small systems \cite{SelEta10a,SelEta13a}. The GK equation was also obtained in the framework on non-equilibrium thermodynamics, assuming a minimal deviation from local equilibrium both in entropy density and in entropy current density \cite{VanFul12a,KovVan15a}. 

It is remarkable that one can obtain the MCV equation when $l=0$, but Fourier-type solutions of the system \re{ebal}--\re{GK} are obtained whenever, with the thermal diffusivity $\alpha = \f{k}{\rho c}$,
\begin{equation}
\f{l^2}{\tau} = \alpha.
\label{Fres}\end{equation}
Hereafter we will call this formula the {\em Fourier resonance condition}. If  $\f{l^2}{\tau}> \alpha$ then the solutions of \re{ebal}--\re{GK} show wavelike characteristics while, if $\f{l^2}{\tau} < \alpha$ then the solutions are over-diffusive \cite{TanAra00a,KovVan15a}. This is due to the hierarchical structure of the system of equations \cite{BerEta11a,EngBer15a}. 

We are looking for solutions of \re{ebal}--\re{GK} with heat pulse boundary condition at the front side. The heat pulse is introduced in the following form:
\begin{center}
	$q_0(t)=q(x=0,t)= \left\{ \begin{array}{cc}
	q_{\rm max}\left(1-\cos\left(2 \pi \cdot \frac{t}{t_p}\right)\right) & 
	\textrm{if }\   0<t \leq t_p,\\
	0 & \textrm{if }\ t>t_p.
	\end{array} \right.  $
\end{center}
This pulse profile is different from the measured one, see Figure \ref{fig:pulse}, but the length of the pulse is much shorter than the characteristic timescale of the experiment so the particular shape is insignificant. For our experiment $t_p = 0.01 s$. The backside boundary is considered adiabatic  $q(L,t)=0$. Initially, the temperature distribution is uniform and the heat flux is zero along the sample, that is, $T(x,0)=T_0 (= 21^{\circ}{\rm C})$ and $q(x,0)=0$. 

It is convenient to introduce a dimensionless form of the equations \re{ebal}--\re{GK}, by the following definitions of the dimensionless variables  $\hat t, \hat x, \hat T, \hat q, \hat Q$ for time, position, temperature, heat flux and the internal variable, respectively: 
\begin{eqnarray}
\hat{t} =\frac{t}{t_p},  \qquad
\hat{x}=\frac{x}{L};\nonumber \\
\hat{q}=\frac{q}{q_{\rm max}}, \quad &\text{where}&\quad
q_{\rm max} = \frac{1}{t_p}  \int_{0}^{t_p} q_{0}(t)dt;\nonumber \\
\hat{T}=\frac{T-T_{0}}{T_{\rm end}-T_{0}}, \quad &\text{where}&\quad
T_{\rm end}=T_{0}+\frac{{q}_{\rm max} t_p}{\rho c L}.\ \ \
\label{ndvar}\end{eqnarray}

The dimensionless parameters are, consequently,
\begin{equation}
\hat{\tau} =\frac{\tau}{t_p}; \quad\ 
\hat{\alpha} 	  = \frac{k t_p}{\rho c L^2}; \quad\  
\hat{l} 	  = \f{l}{L}.
\end{equation}

Accordingly, the nondimensional form of the equations is
\begin{eqnarray}
\f{\partial \hat T}{\partial \hat t} + \f{\partial \hat  q}{\partial \hat  x} &=& 0, \label{ndebal}\\
\hat \tau \f{\partial \hat  q}{\partial \hat  t} + \hat  q + \hat  \alpha \f{\partial \hat  T}{\partial \hat x} - {\hat l}^2 \f{\partial^2 \hat q}{\partial \hat{x}^2} &=& 0. \label{ndGK} 
\end{eqnarray}

The boundary and initial conditions are:

\begin{center}
	$\hat q_0(t)=\hat  q(x=0,t)= \left\{ \begin{array}{cc}
	1-\cos\left(2 \pi \cdot \hat t\right) & 
	\textrm{if } 0<\hat t \leq 1,\\
	0 & \textrm{if } \hat t>1.
	\end{array} \right. $
\end{center}
$\hat  q(1,\hat t)=0$, $\hat T(\hat x,0)=0$ and $\hat q(\hat x,0)=0$. For the Fourier equation, the last initial condition is not necessary.

\section{Solution of the inverse problem}

\begin{figure}[ht]
	\centering
	\includegraphics[width=6cm]{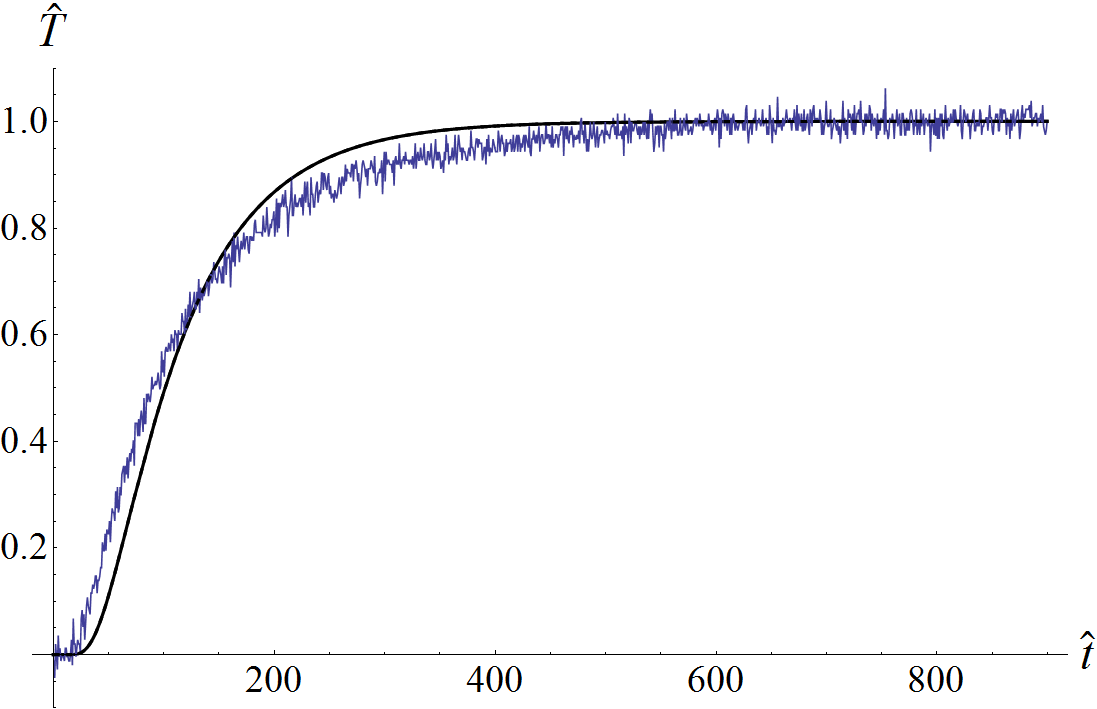}
	\includegraphics[width=6cm]{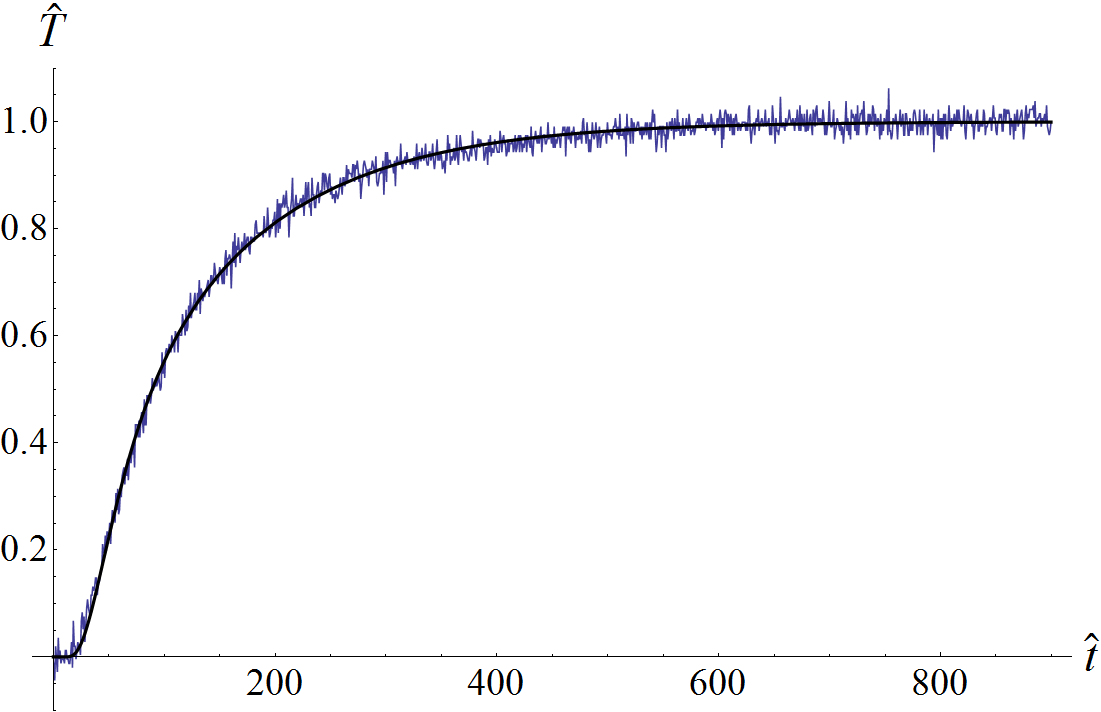}
	\caption{On the left hand side, the best fitted Fourier solution is presented, and on the right hand side the best fitted GK one is visible. The thin noisy line is the experimentally measured data of the backside temperature as the function of time. The solid lines are the solutions of the Fourier equation with thermal diffusivity $\hat \alpha=0.001409\pm 5\cdot 10^{-6}$ on the left hand side, and that of the GK modell \re{ndebal}--\re{ndGK} with $\hat \alpha=0.001288\pm 3\cdot 10^{-6}$, $\hat \tau=50.8\pm 1.2$, and ${\hat l}^2=0.100 \pm 0.002$. }
\label{fig:fit-data}\end{figure}

The system of equations \re{ndebal}--\re{ndGK} is remarkably stable, and this is an advantage for the calculation of the parameters best fitting to a given measurement data. These calculations were performed with the built-in  nonlinear regression algorithm of  Mathematica 8.0, using the solution of the system of partial differential equations as an input function. 
Figure \ref{fig:fit-data} demonstrates our results.

In both figures, the thin, noisy curve is a typical measurement data of a heterogeneous sample, comprising $2250$ data points\footnote{The data set is enclosed to the arXiv publication.}. Both the temperature and the time scales are nondimensional. The solid curves represents the best approximation of the data by the Fourier equation on the left figure and that of the GK equation on the right figure. For the Fourier solution the thermal diffusivity is the only parameter, the regression analysis gives $\alpha = [2.144 \pm 0.008]\cdot 10^{-6} m^2/s$, with $R^2 = 0.9977$. For the GK heat conduction model the best approximation of the data is given by material parameters $\alpha=[1.958\pm 0.004]\cdot 10^{-6} m^2/s$, $\tau=[0.51 \pm 0.01] s$, and $ {l}^2=[1.53 \pm 0.03]\cdot 10^{-6} m^2$. The dimensional standard errors of the statistical analysis are calculated  assuming exact $L$ and $t_p$ values.  The goodness of the regression is better than that of the Fourier one, as it is visible in the figures as well as by the higher  $R^2 = 0.9996$ value.
	
One may observe that the arrival time of the best fit Fourier signal is longer, than the experimentally observed value. The prediction of the MCV equation is a delay compared to Fourier's model. Therefore our experimental observations exclude the MCV equation.

\section{Discussion}

We have presented a heat pulse experiment at room temperature in a macroscopic, heterogeneous specimen, that  cannot be modelled properly either by the Fourier type or the MCV type heat conduction, but can be described by the GK equation in its over-diffusive regime. Over-diffusive GK signals show larger apparent propagation speed compared to the best Fourier approximation, in contrast to the MCV equation.

	
The heterogeneity of the considered specimen is parallel to the heat flux (see Figure \ref{fig:exp}), providing a simple explanation of the phenomenon in terms of coupled heat conductors, where the temperature difference is maintained by the differing conductivities and the respective boundary conditions. In this respect the comparison to the low-temperature experiments can be instructive because there the specific parameter of the GK equation, $l$, is well interpreted. However, according to the universality of the non-equilibrium thermodynamic model, the form of the equation is independent of the particular mechanism. Therefore, the practical applicability of the simple modelling framework and of its established parameters is not restricted to this specific type of heterogeneity. On the other hand, the identification of particular mesomechanisms is extremely important also because then the corresponding material parameters can be calculated and designed.

\section{Acknowledgements}
The work was supported by the grant OTKA  K104260. 

\bibliographystyle{unsrt}

\end{document}